\documentclass[11pt,a4paper]{article}

\usepackage{jcappub}

\usepackage{graphicx}
\usepackage{enumitem}
\usepackage{amsmath}
\usepackage{booktabs}
\usepackage{array}
\usepackage{color}
\usepackage{amssymb}
\usepackage{mathrsfs}
\usepackage[T1]{fontenc}
\usepackage[latin1]{inputenc}
\usepackage[table]{xcolor}  
\usepackage{verbatim}
\usepackage{caption}
\usepackage{subcaption}
\usepackage{rotating}
\usepackage{dcolumn}
\usepackage{bm}

\renewcommand{\d}{\mathrm{d}}

\newcommand{\vect}[1]{\bm{\mathrm{{#1}}}}

\newcommand{\e}[1]{\mathrm{e}^{{#1}}}

\newcommand{\Lie}[1]{\mathcal{L}_{{#1}}}

\newcommand{\ipleft}{\langle\kern-0.2em\langle}
\newcommand{\ipright}{\rangle\kern-0.2em\rangle}

\newcommand{\R}{\mathcal{R}}

\newcommand{\Mp}{M_{\mathrm{P}}}

\renewcommand{\leq}{\leqslant}

\newcommand{\be}{\begin{equation}}
\newcommand{\ee}{\end{equation}}

\DeclareMathOperator{\Or}{O}

\newcommand{\para}[1]{\par\vspace{2mm}\noindent\textbf{{#1}.---}}

\newcolumntype{s}{>{$\displaystyle}l<{$}}
\newcolumntype{t}{>{$\displaystyle}c<{$}}
\newcolumntype{u}{>{$\displaystyle}r<{$}}
\newcolumntype{v}{>{$\displaystyle}m{4cm}<{$}}

\newcolumntype{d}{D{!}{\;\pm\;}{-1}}

\title{The curvature perturbation at second order}
\author{Mafalda Dias,$^1$}
\author{Joseph Elliston,$^1$}
\author{Jonathan Frazer,$^{2,3}$}
\author{\newline David Mulryne$^4$}
\author{and David Seery$^1$}
\affiliation{\small $^1$ Astronomy Centre, University of Sussex,
Falmer, Brighton, BN1 9QH, UK \\
$^2$ Department of Theoretical Physics, University of the Basque Country, UPV/EHU, 48040 Bilbao, Spain\\
$^3$ IKERBASQUE, Basque Foundation for Science, 48011 Bilbao, Spain \\
$^4$ Astronomy Unit, Queen Mary University of London,
Mile End Road, London, E1 4NS, UK}
\abstract{We give an explicit relation,
up to second-order terms,
between scalar-field fluctuations
defined on spatially-flat slices
and the curvature perturbation on uniform-density
slices.
This expression is a necessary ingredient
for calculating observable quantities
at second-order and beyond in multiple-field inflation.
We show that
traditional cosmological perturbation
theory and the `separate universe' approach
yield equivalent expressions
for superhorizon wavenumbers,
and in particular that all nonlocal terms can
be eliminated from the perturbation-theory
expressions.}

\notoc

\begin{document}	
\maketitle

\section{Introduction}
\label{sec:introduction}

According to our current ideas, structure
in the universe was seeded by
quantum fluctuations
which were amplified during an inflationary epoch.
During inflation
these fluctuations dominate the variation in energy density
from place to place
and therefore generate
a gravitational response which can be probed by cosmological observations.

Inflationary amplification is believed to occur for
any sufficiently light degree of freedom,
in the sense that its mass $m$ was substantially less than the
Hubble rate $H$ while scales of interest 
were being carried beyond the horizon.
Models motivated by modern concepts in high-energy physics
often invoke many light fields,
and therefore can be tested only if we
have an understanding
of their effects.
The literature surrounding calculations of the inflationary density
perturbation is now very mature---often with agreement
on subtle effects to second- or even third-order in
perturbation theory---which allows
these effects to be predicted in some detail.
But despite this maturity
it is remarkable that
no completely explicit
formula
has been given for the uniform-density gauge
curvature perturbation in
an inflationary model with an
arbitrary number of fields.%
    \footnote{Maldacena computed the second-order
    version of this relationship
    in a single-field, canonical model of inflation~\cite{Maldacena:2002vr}.
    Some results are known to second-
    or even third-order
    for multiple-field scenarios,
    but typically these invoke
    the slow-roll approximation
    or do not explicitly specialize
    to a scalar field model.
    See, eg. Malik~\cite{Malik:2005cy}.
    Anderson et al. gave results to third order
    for superhorizon scales using the slow-roll
    expansion~\cite{Anderson:2012em}.
    A geometrical description of the large-scale, second-order
    gauge transformation based on curvatures in
    the phase-space of solutions was given in Ref.~\cite{Seery:2012vj}.}
A formula of this type would
give the next-order term in the
classic
result $\zeta = - \dot{\phi}_\alpha \delta \phi^\alpha / 2 \Mp^2 H \epsilon$
which has long been known
at first order.
It is a key element in
computing non-Gaussian
signatures in the statistics of the inflationary density perturbation.
Here and below, $\epsilon \equiv - \dot{H}/H^2$ is the usual slow-roll parameter
and $\delta\phi^\alpha$ labels the species of light fields.

In this paper we supply the missing formula,
valid for an arbitrary number of canonical fields and
without using the slow-roll approximation.
We perform the calculation using two independent
methods:
traditional `cosmological perturbation theory',
which is an expansion in the amplitude
of small fluctuations around a
Robertson--Walker background,
and the `separate universe approach',
which is an expansion in
the amplitude of \emph{gradients}
of the perturbations.
In practice
(although not required in principle),
separate-universe calculations
often invoke a second expansion in the
amplitude of the fluctuations, after which the
two methods should agree for any
Fourier mode much larger than the cosmological horizon.
For a mode of wavenumber $\vect{k}$ this requires $k/aH \ll 1$,
making spatial gradients negligible.
We verify that the two approaches
give equivalent answers
and clarify some issues regarding
nonlocal terms which appear in
the perturbation theory expressions.

Our final expressions will be used in forthcoming
papers
which describe
numerical calculation of the two- and three-point
functions in multiple-field inflation.

While the final version of this paper was being
prepared, a preprint by Christopherson,
Nalson \& Malik appeared in which the
second-order gauge transformation was
given explicitly for a scalar field model~\cite{Christopherson:2014bea}.
We comment on the relation between our results
in~{\S\ref{sec:conclusions}}.

\para{Notation}%
We work in units where $c=\hbar=1$.
Newton's constant is expressed in terms of the
reduced Planck mass, $\Mp^2 = (8\pi G)^{-1}$.
Spacetime indices are labelled with Greek
letters from the middle of the alphabet, $\mu$, $\nu$, \ldots,
and spatial indices
are labelled with Latin indices from the
middle of the alphabet, $i$, $j$, \ldots.
The different species of scalar field are
labelled with Greek letters from the beginning of the
alphabet, $\alpha$, $\beta$, \ldots.

\section{Cosmological gauges}
\label{sec:gauges}

The unperturbed cosmology is taken to be
described by a
spatially flat Robertson--Walker metric
\begin{equation}
    \d s^2 = - \d t^2 + a(t)^2 \d \vect{x}^2 ,
\end{equation}
where $a(t)$ is the scale factor and $H = \dot{a}/a$ is the Hubble
parameter. An overdot denotes a derivative with respect to cosmic
time $t$.

\para{Choice of slicing}%
In the unperturbed universe,
spatial hypersurfaces of fixed
time $t$ are associated with a number of physical
properties: they are slices of uniform energy density,
uniform Hubble parameter,
zero intrinsic Ricci curvature, and so on.
Once we add perturbations
these
hypersurfaces continue to exist
but typically no longer coincide.
To compare the value of some physical quantity such as 
the density $\rho$
between the perturbed and unperturbed universes
we pick one set of hypersurfaces to use as a reference.
This is said to be a \emph{choice of slicing}.
The perturbation in a physical quantity is
defined to be the difference between its value on the
same hypersurface
in the perturbed and unperturbed universes.

A choice of slicing, together with a rule for determining the
spatial coordinates on each slice, is called a choice of gauge.
In principle we can fix the slicing and use whatever coordinate system we like to
describe it, but in practice it is convenient to choose coordinates so that slices
of constant $t$ coincide with the slicing.
We describe coordinates with this property as \emph{adapted to the slicing}.
Having chosen a slicing,
the metric can be written in adapted
coordinates using Arnowitt--Deser--Misner (ADM) quantities,
\begin{equation}
    \d s^2 = - N^2 \, \d t^2 +  h_{ij} ( \d x^i + N^i \d t) ( \d x^j + N^j \d t ) ,
\end{equation}
where $N$ is the lapse function and $N_i$ the shift vector.
The spatial metric $h_{ij}$ is used to raise and lower
spatial indices, eg. $N_i = h_{ij} N^j$.
The curvature perturbation associated with this
slicing, denoted $\psi$, is defined by%
    \footnote{This definition is not the same as that of
    the review article by
    Malik \& Wands~\cite{Malik:2008im}.
    It agrees with the quantity
    used at the nonlinear level by
    Maldacena~\cite{Maldacena:2002vr}.
    The definition~\eqref{eq:psi-def}
    was used
    to prove conservation of $\psi = \zeta$
    in the uniform density gauge
    at a classical level
    by
    Shellard \& Rigopoulos~\cite{Rigopoulos:2003ak},
    Lyth, Malik \& Sasaki~\cite{Lyth:2004gb}
    and Weinberg~\cite{Weinberg:2008nf,Weinberg:2008si}.
    More recently the proof has been
    strengthened to
    an operator statement in
    quantum mechanics by
    Assassi, Baumann \& Green~\cite{Assassi:2012et}.}
\begin{equation}
    \e{6\psi} \equiv \det (h_{ij}/a^2) .
    \label{eq:psi-def}
\end{equation}
A number of slicings are commonly used in the literature.
The most important are:
\begin{itemize}
\item \textsf{Spatially flat slicing.}
This has
$\det (h_{ij}/a^2) = 1$
and therefore $\psi$ is identically zero.
In the absence of gravitational waves
there exist coordinates for which
$h_{ij} = a^2 \delta_{ij}$.
The Ricci curvature of each spatial hypersurface is zero.

If gravitational waves are present then
$h_{ij} = a^2 \e{\gamma_{ij}}$ where $\gamma_{ij}$ is transverse and
traceless. This preserves the condition $\det (h_{ij}/a^2)  = 1$
but the Ricci curvature is no longer zero.
In this
context
we should more properly speak of a
`uniform Hubble slicing'.

\item \textsf{Comoving slicing.}
This is chosen so that there is no net energy flux measured on a fixed
slice.
Applied to the energy--momentum tensor
in a holonomic basis adapted to the slicing this implies
$T_{0i} = 0$.
The curvature perturbation defined by this slicing is
conventionally denoted $\R$.%
    \footnote{There are differing sign conventions for $\R$. Our definition
    gives $\zeta = \R + \Or(k/aH)^2$ on superhorizon scales,
    but other definitions reverse this
    to $\zeta = -\R + \Or(k/aH)^2$.}

\item \textsf{Uniform density slicing.}
The density $\rho$ is constant on a fixed slice.
The curvature perturbation is
conventionally written $\zeta$.
In the absence of gravitational waves, 
coordinates exist in which the
spatial metric can be written $h_{ij} = a^2 \e{2 \zeta} \delta_{ij}$.
\end{itemize}

It is known that $\R$ and $\zeta$ agree on superhorizon scales
up to second order,
in the sense that $(\R - \zeta)_{\leq 2} = \Or(k/aH)^2$~\cite{Malik:2008im},
where the subscript `$\leq2$' denotes terms of second order or less.
We will reproduce the first-order version
of this result by direct calculation
in~\S\ref{sec:uniform-density-zeta} below.
In this paper we focus on
$\zeta$ because it is known to be conserved
to all orders in perturbation theory
(including quantum effects)
when the dynamics are
adiabatic~\cite{Malik:2003mv,Rigopoulos:2003ak,Lyth:2004gb,Weinberg:2008nf,Weinberg:2008si,Assassi:2012et}.
To our knowledge the
equivalence between $\zeta$ and $\R$,
and conservation of $\R$ in an adiabatic regime,
have been explicitly demonstrated only
to second order~\cite{Vernizzi:2004nc}.

In the absence of isocurvature perturbations,
$\zeta$
can be used to set initial conditions for the 
CMB anisotropy.
Therefore it represents
a convenient way to express observable quantities.
But
inflationary calculations
are often technically simplest
in the spatially flat gauge,
where the curvature perturbation is zero
and fluctuations are measured by the
scalar field perturbations
$\delta\phi^\alpha$.
If we take advantage of this simplicity then a rule
is needed to connect the
$\delta\phi^\alpha$ to $\zeta$.
As explained in \S\ref{sec:introduction},
our objective is to compute this rule to
second order in the $\delta\phi^\alpha$.

This approach was used by
Guth \& Pi~\cite{Guth:1982ec}
and Bardeen, Steinhardt \& Turner~\cite{Bardeen:1983qw}
in the earliest
estimates of the density perturbation.
These calculations 
exploited the technical simplicity of the flat gauge
to compute the amplification of quantum effects,
after which a variety of arguments
were used to estimate
the first-order, single-field result
$\zeta \sim -H \delta{\phi}/\dot{\phi}$~\cite{Guth:1982ec}.
The relation between these methods
was clarified by Lyth~\cite{Lyth:1984gv}.
Later, the first-order result was extended to multiple-field
scenarios by Salopek \& Bond~\cite{Salopek:1988qh},
who used it to generate numerical results.
Formulae for more complex models were given
by Sasaki \& Stewart
using
the `separate universe
approach'~\cite{Starobinsky:1986fxa,Sasaki:1995aw,Sasaki:1998ug}.
More recently,
Maldacena computed the relationship between
$\zeta$ and $\delta\phi$ in a single-field
model
and discussed its application to higher $n$-point functions~\cite{Maldacena:2002vr}.

\subsection{Changing slicing}
\label{sec:changing-slice}

To connect quantities defined by different slicings, such as
$\zeta$ and $\delta\phi^\alpha$,
we must change the gauge.
In the literature
this is sometimes described as a coordinate transformation.
If not interpreted correctly this description is confusing because
under a coordinate transformation any tensor
transforms covariantly,
and we shall see that this is not the same as the transformation
law under a change of gauge.
The difference arises because to change gauge we
first change the slicing
and \emph{then} change the coordinates to adapt to it.

Begin with some initial slicing and adapted coordinates $x^\mu$.
Suppose we wish to switch to a different set of slices
which are slightly displaced.
At any point $p$ the displacement to the matching point $p'$
on the new surface is written
\begin{equation}
    x^\mu(p) \rightarrow x^\mu(p') = \e{\Lie{\xi}} x^\mu(p) .
    \label{eq:coordinate-map}
\end{equation}
The Lie derivative $\Lie{\xi}$ is understood to act on the
coordinates $x^\mu(p)$ as if they were the components of a
contravariant vector field.
This abuse of notation is unfortunate but conventional.
The vector $\xi^\mu$ associated with the Lie derivative
is called the gauge parameter.
Given two slicings our task will usually be to solve for
an appropriate gauge parameter.

Now introduce a second set of coordinates $x^{\bar{\mu}}$ adapted to the
new slicing,
with the time coordinate adjusted so that the numerical
value of time agrees on both slices.
We distinguish indices associated with these new coordinates using a bar.
By a `gauge transformation', we mean a map from
tensors at $p$ expressed in
the holonomic basis
basis $\{ \d x^\mu, \partial / \partial x^\mu \}$
to tensors at $p'$ expressed in
the holonomic
basis $\{ \d x^{\bar{\mu}}, \partial / \partial x^{\bar{\mu}} \}$.
This is both a change of evaluation point and
a change of basis.

To define the map, consider a generic tensor
$\vect{T}$ at $p$.
In the original basis its components are
\begin{equation}
	\vect{T}|_p = {T_{\mu \cdots}}^{\nu \cdots}(p) \,
	\d x^\mu|_p \otimes \cdots \otimes
	\left.\frac{\partial}{\partial x^\nu}\right|_p \otimes \cdots.
	\label{eq:T-def}
\end{equation}
The gauge transformation yields a transformed tensor
$\vect{T}'$ at $p'$,
\begin{equation}
	\vect{T}'|_{p'} = {T_{\bar{\mu} \cdots}}^{\bar{\nu} \cdots}(p') \,
	\d x^{\bar{\mu}}|_{p'} \otimes \cdots \otimes
	\left.\frac{\partial}{\partial x^{\bar{\nu}}}\right|_{p'} \otimes \cdots .
	\label{eq:Tprime-def}
\end{equation}
The map between the components, expressed in these different
bases, is
\begin{equation}
    {T_{\mu \cdots}}^{\nu \cdots}(p) \rightarrow
    {T_{\bar{\mu} \cdots}}^{\bar{\nu} \cdots}(p')
    =
    \e{\Lie{\xi}}
    {T_{\mu \cdots}}^{\nu \cdots}(p)
    \label{eq:gauge-xfm}
\end{equation}
where one should identify
matching index labels $\mu \rightarrow \bar{\mu}$
(and so on), and
on the right-hand side the Lie derivative
is understood to mean its action on the components
of $\vect{T}$ in the original basis.%
    \footnote{Note that this map must be phrased carefully.
    In the literature of `active' gauge transformations,
    which is the point of view being adopted here,
    one sometimes finds the statement
    \begin{equation}
    	\vect{T} \rightarrow \vect{T}' = \e{\Lie{\xi}} \vect{T}.	
    \end{equation}
	To see that this can be made to agree with our
	Eq.~\eqref{eq:gauge-xfm} requires an extra assumption.
    The action of the Lie derivative on
    the tensor $\vect{T}$ defined in~\eqref{eq:T-def}
    yields another tensor at $p$,
    \begin{equation}
    	\Lie{\xi}\vect{T}
    	=
    	\Lie{\xi}
    	\Big(
    		{T_{\mu \cdots}}^{\nu \cdots}(p) \,
    		\d x^\mu|_p \otimes \cdots \otimes
    		\left.\frac{\partial}{\partial x^\nu}\right|_p \otimes \cdots
    	\Big)
    	=
    	\big[
    		\Lie{\xi} {T_{\mu \cdots}}^{\nu \cdots}
    	\big]_p
		\d x^\mu|_p \otimes \cdots \otimes
   		\left.\frac{\partial}{\partial x^\nu}\right|_p \otimes \cdots .
   		\label{eq:lie-action}
    \end{equation}
    The notation $\Lie{\xi} {T_{\mu \cdots}}^{\nu \cdots}$ denotes
    the action of the Lie derivative in a coordinate basis; for example,
    for a one-form or covariant vector,
    $\Lie{\xi} \omega_a = \xi^b \partial_b \omega_a + \omega_b \partial_a x^b$.
    To obtain a tensor evaluated at $p'$ from~\eqref{eq:lie-action}
    requires a separate rule,
    which is the extra assumption described above.
    
    One option is to use the push-forward or Jacobian map,
    which would undo part of the action of the Lie derivative.
    This is defined using the Jacobian map to pull back the basis at
    $p'$ to $p$,
    so after doing so the components would be related only
    by a change of evaluation point,
    \begin{equation}
    	\Lie{\xi}\vect{T} \xrightarrow{\text{Jacobian map}}
    	{T_{\mu \cdots}}^{\nu \cdots}(p') \,
    	\d x^\mu|_{p'} \otimes \cdots \otimes
    	\left.\frac{\partial}{\partial x^{\nu}}\right|_{p'} \otimes \cdots .
    \end{equation}
	This reproduces the starting point for the gauge transformation,
	described above Eq.~\eqref{eq:T-def}.
	Therefore, after changing basis
	to $(\d x^{\bar{\mu}})_{p'}$, $(\partial/\partial x^{\bar{\mu}})_{p'}$
	and exponentiating the Lie derivative operator,
	one will again arrive at Eq.~\eqref{eq:gauge-xfm}.}

In the context of perturbation theory, the displacement
between hypersurfaces is small and therefore
so is the gauge parameter $\xi^\mu$.
In this paper we are interested in computing the relationship
between quantities defined on different slicings up to
second order in amplitude.
Hence, we must work to the same order in powers of
$\xi^\mu$.
We break $\xi^\mu$ into temporal and spatial gauge parameters
$\xi^0$ and $\xi^j$,
corresponding to the time and space components of $\xi^\mu$.

\para{Transformation of field fluctuations}%
Using Eq.~\eqref{eq:gauge-xfm}
we can compute how each quantity of interest
transforms between slicings.
A field fluctuation $\delta\phi^\alpha$ transforms according to the rule
\begin{equation}
    \delta\phi^\alpha(t')
    =
        \delta\phi^\alpha(t) +
        \xi^0 \dot{\phi}^\alpha +
        \xi^0 \frac{\partial \delta\phi^\alpha(t)}{\partial t}
        + \xi^j \partial_j \delta\phi^\alpha(t)
        + \frac{(\xi^0)^2}{2} \ddot{\phi}^\alpha
        + \frac{\dot{\phi}^\alpha}{4}
            \frac{\partial(\xi^0)^2}{\partial t} .
\end{equation}
On the right-hand side, $\dot{\phi}^\alpha$, $\ddot{\phi}^\alpha$
(and so on)
represent derivatives of the background field with respect to time.
Because we adjusted the time coordinates of the slices to agree
it is not necessary to specify whether the derivatives are with respect
to $t$ or $t'$.
The symbols $\delta\phi^\alpha(t)$
and $\delta\phi^\alpha(t')$ denote,
respectively,
field fluctuations defined on the first slicing of constant $t$,
and the second slicing of constant $t'$.

The time derivative of a field fluctuation transforms according to
\begin{equation}
\begin{split}
    \frac{\partial \delta\phi^\alpha(t')}{\partial t'}
    = \mbox{} &
        \frac{\partial \delta\phi^\alpha(t)}{\partial t}
        + \xi^0 \frac{\partial^2 \delta\phi^\alpha(t)}{\partial t^2}
        + \xi^j \partial_j \frac{\partial \delta\phi^\alpha(t)}{\partial t}
        + \frac{\partial \xi^0}{\partial t} \frac{\partial \delta\phi^\alpha(t)}{\partial t}
        + \frac{\partial \xi^j}{\partial t} \partial_j \delta\phi^\alpha(t)
    \\ & \mbox{}
        + \ddot{\phi}^\alpha \xi^0
        + \frac{1}{2} \dddot{\phi}^\alpha (\xi^0)^2
        + \dot{\phi}^\alpha \frac{\partial \xi^0}{\partial t}
        + \frac{3}{4} \ddot{\phi}^\alpha \frac{\partial (\xi^0)^2}{\partial t}
        + \frac{\ddot{\phi}^\alpha}{2} \xi^j \partial_j \xi^0
    \\ & \mbox{}
        + \frac{\dot{\phi}^\alpha}{4} \frac{\partial^2(\xi^0)^2}{\partial t^2}
        + \frac{\dot{\phi}^\alpha}{2} \frac{\partial}{\partial t}
            \Big(
                \xi^j \partial_j \xi^0
            \Big) .
\end{split}
\end{equation}

\para{Transformation of metric components}%
We also require transformation rules for the metric components
$N$, $N^i$ and $h_{ij}$.
Bearing in mind that we intend to compute $\zeta$ in terms
of the flat-gauge perturbations $\delta\phi^\alpha$
we
simplify these expressions by assuming
that the initial slicing corresponds
to the flat gauge where $h_{ij} = a^2 \delta_{ij}$.
We do not yet impose any restriction on the final slicing.

Instead of working with the lapse directly
it is more convenient to work in terms of its
perturbation $\alpha$, defined by $N \equiv 1 + \alpha$.
The transformation rule for $\alpha$ is%
    \footnote{Contraction of repeated indices in the lowered position
    implies summation with the Euclidean metric $\delta_{ij}$.}
\begin{equation}
\begin{split}
    \alpha'
    = \mbox{} &
        \alpha
        + \frac{\partial \xi^0}{\partial t}
        + \frac{1}{4} \frac{\partial^2 (\xi^0)^2}{\partial t^2}
        + \frac{1}{2} \frac{\partial}{\partial t}
            \Big(
                \xi^j \partial_j \xi^0
            \Big)
        + \frac{\partial (\alpha \xi^0)}{\partial t}
        + \xi^j \partial_j \alpha
        - N^m \partial_m \xi^0
    \\ & \mbox{}
        + \frac{1}{2a^2}\partial_i \xi^0 \partial_i \xi^0
        - \partial_i \xi^0 \frac{\partial \xi^i}{\partial t} .
\end{split}
\end{equation}
Likewise, the transformation rule for the shift vector is
\begin{equation}
\begin{split}
    N_{j'}
    = \mbox{} &
        N_j
        - \partial_j \xi^0
        - \frac{1}{4} \partial_j \frac{\partial(\xi^0)^2}{\partial t}
        - \frac{1}{2} \partial_j (\xi^m \partial_m \xi^0)
        - \frac{\partial \xi^0}{\partial t} \partial_j \xi^0
        - 2 \alpha \partial_j \xi^0 
        + N_j \frac{\partial \xi^0}{\partial t}
    \\ & \mbox{}
        + N_m \partial_j \xi^m
        + \xi^0 \frac{\partial N_j}{\partial t}
        + \xi^m \partial_m N_j
    \\ & \mbox{}
        + a^2\Big(
            \delta_{jm} (1 + 2 H \xi^0) \frac{\partial \xi^m}{\partial t}
            + \delta_{mn} \frac{\partial \xi^m}{\partial t} \partial_j \xi^n
            + \frac{\delta_{jm}}{2} \frac{\partial}{\partial t}
            \Big[
                \xi^0 \frac{\partial \xi^m}{\partial t}
                + \xi^n \partial_n \xi^m
            \Big]
        \Big) .
\end{split}
\label{eq:shift-xfm}
\end{equation}
Finally, the spatial metric transforms according to
\begin{equation}
\begin{split}
    h_{i'j'} = \mbox{} &
        a^2 (1 + 2H \xi^0) \delta_{ij}
        - \partial_i \xi^0 \partial_j \xi^0
        + N_i \partial_j \xi^0
        + N_j \partial_i \xi^0
    \\ & \mbox{}
        + a^2 \Big(
            (1 + 2H\xi^0) (\delta_{im} \partial_j + \delta_{jm} \partial_i) \xi^m
            + \Big[
                    \frac{\delta_{im}}{2} \partial_j
                    + \frac{\delta_{jm}}{2} \partial_i
                \Big]
                \Big[
                    \xi^0 \frac{\partial \xi^m}{\partial t} + \xi^n \partial_n \xi^m
                \Big]
    \\ & \mbox{}
        \qquad\quad
            + \delta_{mn} \partial_i \xi^m \partial_j \xi^n
            + \delta_{ij} (2H + \dot{H})(\xi^0)^2
            + \frac{\delta_{ij}}{2} H \frac{\partial(\xi^0)^2}{\partial t}
        \Big) .
\end{split}
\label{eq:spatial-metric-xfm}
\end{equation}
From~\eqref{eq:psi-def}
and~\eqref{eq:spatial-metric-xfm}
we can compute the curvature
perturbation in the new slicing. It is
\begin{equation}
\begin{split}
    \psi' = \mbox{} &
        H \xi^0
        + \frac{1}{3} \partial_j \xi^j
        - \frac{1}{6a^2} \partial_j \xi^0 \partial_j \xi^0
        + \frac{1}{3} N^m \partial_m \xi^0
        + \frac{1}{6} \partial_j \xi^0 \frac{\partial \xi^j}{\partial t}
        + \frac{1}{6} \xi^0 \frac{\partial (\partial_j \xi^j)}{\partial t}
    \\ & \mbox{}
        + \frac{1}{6} \xi^m \partial_m \partial_j \xi^j
        + \frac{H}{4} \frac{\partial(\xi^0)^2}{\partial t}
        + \frac{\dot{H}}{2} (\xi^0)^2 .
\end{split}
\label{eq:psi-xfm}
\end{equation}

The definition $\psi \sim \det h/a^2$
implies that the curvature perturbation measures modulation
in proper volume from place to place on a fixed slice.
Eq.~\eqref{eq:psi-xfm}
exhibits the expected invariance
under volume-preserving
transformations of the spatial coordinates
which do not change the slicing.
These are generated by gauge transformations with
$\xi^0 = 0$ and divergenceless $\xi^j$, \emph{viz.}
$\partial_j \xi^j = 0$.
They include the spatial rotations.

Gauge transformations with $\xi^0 \neq 0$
change the slicing.
For such transformations
there is a small second-order volume modulation
even if $\xi^j$ is divergenceless,
provided it is time-dependent
and
$\xi^0$ is spatially dependent.
This arises from the second-to-last
term in the first line of~\eqref{eq:psi-xfm}.
If $\xi^j$ is time-independent there is no
modulation, and no contribution to the
curvature perturbation.
Eq.~\eqref{eq:shift-xfm}
shows that a time-independent transformation
of this kind
negligibly perturbs the shift-vector
$N^j$
when all $\vect{k}$-modes are associated with
superhorizon scales
for which $k / aH \ll 1$.%
    \footnote{Our interest lies in using the
    second-order gauge transformation to compute
    three- and higher $n$-point correlation
    functions of $\zeta$.
    For this purpose we need an expression such
    as~\eqref{eq:shift-xfm}
    only in the case where each $\xi^0(\vect{k})$ mode
    individually satisfies $k/aH \ll 1$,
    making decay obvious term-by-term.
    A more general theorem
    was proved by Weinberg~\cite{Weinberg:2008nf,Weinberg:2008si}.}

\para{Restriction to diagonal metric}%
Normally only $\xi^0$ is needed to select
the slicing of interest, leaving $\xi^j$ undetermined.
As described above, this ambiguity is irrelevant if
$\xi^j$ becomes time-independent and volume-preserving
when all modes are superhorizon.
More generally we could choose $\xi^j$ to bring
$h_{i'j'}$ to a diagonal form.
This requires
the first-order perturbation to satisfy
$\partial_i \xi^j_1 = 0$,
which
forces $\xi^j_1$ to be spatially homogeneous
(but perhaps time-dependent)
and therefore volume-preserving.
At second order the diagonal constraint is more complex,
but entails 
\begin{equation}
    \partial_j \xi^j_2 =
        \frac{1}{2a^2} \partial_j \xi^0 \partial_j \xi^0
        - N^j \partial_j \xi^0
        - \frac{1}{2} \partial_j \xi^0 \frac{\partial \xi^j}{\partial t} .
    \label{eq:diagonal-condition}
\end{equation}
When $\xi^j$ is chosen to satisfy~\eqref{eq:diagonal-condition}
it can be checked that $\psi'$ becomes independent of
its precise value.
We find
\begin{equation}
    \psi' \overset{\text{diagonal}}{=}
        H \xi^0
        + \frac{H}{4} \frac{\partial (\xi^0)^2}{\partial t}
        + \frac{\dot{H}}{2} (\xi^0)^2 .
    \label{eq:psi-simple}
\end{equation}

The right-hand side of~\eqref{eq:diagonal-condition}
decays when all wavenumbers are
associated with superhorizon scales.
Therefore, on these scales, any rigid volume-preserving
spatial gauge transformation leaves
$h_{i'j'}$ diagonal
and allows $\psi'$ to be computed using
the simplified expression~\eqref{eq:psi-simple}.
Conversely,
because different possibilities for $\xi^j$
change $\psi'$ when $k/aH \gtrsim 1$
there is no unique value of the
curvature perturbation
associated with subhorizon scales.
In practice this is harmless because on these
scales $\psi'$ has no clear significance.

\subsection{Spatially flat slicing}
\label{sec:spatially-flat-slice}

Now we apply this formalism to translate between the
spatially flat slicing and the uniform-density slicing.
In the language of~{\S\ref{sec:changing-slice}},
slices of constant $t$ correspond to the flat gauge
and slices of constant $t'$ correspond to the
uniform density gauge.
The transformed curvature perturbation $\psi'$
will be $\zeta$.

We begin from coordinates in which the flat-gauge
spatial metric is diagonal, \emph{viz.} $h_{ij} = a^2 \delta_{ij}$.
We choose $\xi^0$ to select an appropriate final slicing
and assume that the spatial gauge transformation is chosen
to satisfy~\eqref{eq:diagonal-condition}.

\para{Lapse and shift}%
Before embarking on the calculation, we use this section to collect
formulae for the lapse and shift in the spatially flat gauge.
Eq.~\eqref{eq:psi-xfm} shows that these are not directly
required to compute $\zeta$---%
this expression does not contain $\alpha$, and its $N^j$ dependence
drops out when all wavenumbers are associated with superhorizon scales.
However, they are required indirectly because the density perturbation
which will be used to determine $\xi^0$ depends on the metric.
Moreover, the lapse and shift are elements in an
important constraint equation---the Hamiltonian constraint---which
we will use later to simplify our results.

We work perturbatively in the scalar field fluctuation
$\delta\phi^\alpha$.
We break the shift vector $N_j$ into irrotational and solenoidal
components $\vartheta$ and $\beta$,
\begin{equation}
    N_j \equiv \partial_j \vartheta + \beta_j
\end{equation}
where $\partial_j \beta_j = 0$.
Then $\vartheta$, $\beta_j$ and the lapse perturbation $\alpha$
can be expanded
in powers of $\delta\phi^\alpha$, giving
\begin{equation}
    \alpha \equiv \sum_{n=1}^\infty \alpha_n,
    \quad
    \vartheta \equiv \sum_{n=1}^\infty \vartheta_n
    \quad
    \text{and}
    \quad
    \beta_j \equiv \sum_{n=1}^\infty \beta_{n|j}
\end{equation}
where the term $\alpha_n$ contains exactly $n$ factors of $\delta\phi$,
and likewise for $\vartheta_n$ and $\beta_{n|j}$.

We neglect tensor perturbations, which correspond to
gravitational waves. These could be kept but because they are
represented by transverse traceless tensors $\gamma_{ij}$
and are uncorrelated with the field fluctuations at tree-level
they do not enter connected tree-level autocorrelation functions
of $\zeta$
lower than the trispectrum.
With these choices the lapse perturbations
satisfy~\cite{Maldacena:2002vr,Seery:2005gb,Seery:2006tq}
\begin{subequations}
\begin{align}
    \alpha_1 & = \frac{\dot{\phi}_\alpha \delta \phi^\alpha}{2\Mp^2 H}
    \label{eq:lapse-one}
    \\
    \alpha_2 & = \frac{\alpha_1^2}{2} + \frac{\partial^{-2}}{2H\Mp^2}
        \Big(
            \partial_j \delta\dot{\phi}^\alpha \partial_j \delta\phi_\alpha
            + \delta\dot{\phi}^\alpha \partial^2 \delta\phi_\alpha
            + \frac{1}{a^2} \partial^2 \alpha_1 \partial^2 \vartheta_1
            - \frac{1}{a^2} \partial_i \partial_j \alpha_1 \partial_i \partial_j \vartheta_1
        \Big) .
    \label{eq:lapse-two}
\end{align}
\end{subequations}
This expression for $\alpha_2$ already signals a potential difficulty
because it involves the nonlocal inverse Laplacian $\partial^{-2}$,
defined as multiplication by $-1/k^2$ in Fourier space.
Terms of this nature cannot arise in the separate universe approach
because
it corresponds to an expansion in purely positive powers of $k$.
To demonstrate that a perturbation-theory expression involving
such terms is compatible with a separate-universe calculation we must
show carefully how all nonlocal pieces disappear from the result.
We will do this explicitly in~\S\ref{sec:uniform-density-zeta}.

The first-order component of the scalar shift
satisfies~\cite{Maldacena:2002vr,Seery:2005gb}
\begin{subequations}
\begin{equation}
    - \frac{4H}{a^2} \Mp^2 \partial^2 \vartheta_1 =
        2 V_\alpha \delta\phi^\alpha
        + 2 \dot{\phi}^\alpha \delta\dot{\phi}_\alpha
        + 2 \alpha_1 (6H^2 \Mp^2 - \dot{\phi}^2   ) ,
    \label{eq:hamiltonian-first}
\end{equation}
where $\dot{\phi}^2 \equiv \dot{\phi}^\alpha \dot{\phi}_\alpha$
and
$V_\alpha \equiv \partial_\alpha V$
(and likewise for higher derivatives).
At second order we have~\cite{Seery:2006tq}
\begin{equation}
\begin{split}
    - \frac{4H}{a^2} \Mp^2 \partial^2 \vartheta_2 = \mbox{} &
        \frac{1}{a^2} \partial_j \delta\phi^\alpha \partial_j \delta\phi_\alpha
        + V_{\alpha\beta} \delta\phi^\alpha \delta\phi^\beta
        + \delta\dot{\phi}^\alpha \delta\dot{\phi}_\alpha
        - \frac{2}{a^2} \dot{\phi}^\alpha \partial_j \vartheta_1 \partial_j \delta\phi_\alpha
    \\ & \mbox{}
        - \frac{\Mp^2}{a^4} \partial^2 \vartheta_1 \partial^2 \vartheta_1        
        + \frac{\Mp^2}{a^4} \partial_i \partial_j \vartheta_1 \partial_i \partial_j \vartheta_1
        + 2H^2 \Mp^2(2\alpha_2 - 3 \alpha_1^2)(\epsilon-3)
    \\ & \mbox{}
        - 2 \alpha_1
            \Big(
                \frac{4 H}{a^2} \Mp^2 \partial^2 \vartheta_1
                + 2 \dot{\phi}^\alpha \delta\dot{\phi}_\alpha
            \Big) .
\end{split}
\label{eq:hamiltonian-second}
\end{equation}
\end{subequations}
At linear order $\beta_{1|j} = 0$.
The second-order component $\beta_{2|j}$ can appear in scalar
quantities only at third order or above because it is divergenceless,
and
therefore is not needed.

\para{Hamiltonian constraint}%
Eqs.~\eqref{eq:hamiltonian-first}--\eqref{eq:hamiltonian-second}
are the first- and second-order parts of the `Hamiltonian constraint',
so called because in Einstein gravity
it is enforced by the lapse $N$ acting as its Lagrange multiplier.
Because the lapse is associated with time reparametrization
invariance the Hamiltonian constraint plays a role analogous to the
Hamiltonian in conventional theory.

We are primarily interested in the case where all $\vect{k}$-modes
are associated with superhorizon scales.
In this limit, $\partial^2 \vartheta_n/a^2$
decays~\cite{Weinberg:2008nf,Weinberg:2008si,Sugiyama:2012tj}
and the Hamiltonian constraint becomes
\begin{equation}
    V_\alpha \delta\phi^\alpha
    + \frac{1}{2} V_{\alpha\beta} \delta\phi^\alpha \delta\phi^\beta
    + \dot{\phi}^\alpha \delta\dot{\phi}_\alpha
    + \frac{1}{2} \delta\dot{\phi}^\alpha \delta\dot{\phi}_\alpha
    + H^2 \Mp^2 (2\alpha_1+2\alpha_2 - 3\alpha_1^2)(3-\epsilon)
    - 2 \alpha_1 \dot{\phi}^\alpha \delta\dot{\phi}_\alpha
    = 0 .
    \label{eq:hamiltonian-constraint}
\end{equation}

\subsection{The uniform-density curvature perturbation}
\label{sec:uniform-density-zeta}

In this section we compute the gauge transformation parameter
$\xi^0$.
To simplify the calculation we take $\xi^j = 0$ from the outset.
On superhorizon scales this will satisfy~\eqref{eq:diagonal-condition},
giving a diagonal spatial metric and trivial lapse.
In~{\S\ref{sec:separate-universe}} we will see that this statement
(promoted to all orders in perturbation theory)
is the basis of the separate universe approach.

\para{Density perturbation}%
Each slicing defines a field of normal vectors $n^\mu$
which are orthogonal to the slices. We normalize so that $n^\mu n_\mu = -1$.
The density measured by an observer
on a fixed spatial slice is $\rho = T_{\mu\nu} n^\mu n^\nu$,
where $T_{\mu\nu}$ is the energy--momentum tensor.
In a holonomic basis of coordinates adapted to the slicing,
this gives
\begin{equation}
    \rho = - \frac{T^{00}}{g^{00}} .
    \label{eq:rho-def}
\end{equation}
Therefore, up to second order, the perturbation in the density
will be
\begin{equation}
    \delta \rho =
        \delta T^{00}
        + \rho \delta g^{00}
        + ( \delta T^{00} + \rho \delta g^{00} ) \delta g^{00} .
    \label{eq:delta-rho}
\end{equation}
Eqs.~\eqref{eq:rho-def} and~\eqref{eq:delta-rho}
apply for any slicing.
Our interest lies in the uniform-density slicing,
for which the density perturbation $\delta\rho(t')$
on slices of constant $t'$ is identically zero.
Using the gauge-transformation formulae collected
in~{\S\ref{sec:changing-slice}}
it is possible to express $\delta\rho(t')$ in terms
of quantities defined on the original flat slices of constant $t$.
That gives
\begin{equation}
    \delta\rho(t') =
        \delta\rho(t)
        + \dot{\rho} \xi^0
        + \delta\dot{\rho} \xi^0
        + \frac{\dot{\rho}}{2} \xi^0 \dot{\xi}^0
        + \frac{\ddot{\rho}}{2} (\xi^0)^2
        - 2
        \Big(
            \delta T^{0i}(t)
            + \rho \delta g^{0i}(t)
        \Big) \partial_i \xi^0
        + \partial_i \xi^0 \partial_i \xi^0
        \Big(
            T^{ij} + \rho g^{ij}
        \Big) .
    \label{eq:gauge-condition}
\end{equation}
The combination $T^{ij} + \rho g^{ij}$ in the final bracket
depends only on background quantities.
Setting the left-hand side equal to zero,
Eq.~\eqref{eq:gauge-condition}
represents an equation for the gauge parameter $\xi^0$ which can be
solved to find the transformation between flat and
uniform-density slices.

\para{Curvature perturbation}%
The solution is
\begin{equation}
\begin{split}
    \xi^0 = \mbox{} &
        - \frac{\delta\rho}{\dot{\rho}}
        + \frac{\delta\dot{\rho}}{\dot{\rho}}\frac{\delta\rho}{\dot{\rho}}
        - \frac{1}{2} \frac{\delta\rho}{\dot{\rho}}
            \frac{\partial}{\partial t} \frac{\delta\rho}{\dot{\rho}}
        - \frac{1}{2} \frac{\ddot{\rho}}{\dot{\rho}}
            \Big( \frac{\delta\rho}{\dot{\rho}} \Big)^2
        + \frac{2}{\dot{\rho}} (\delta T^{0i} + \rho \delta g^{0i})
            \frac{\partial_i \delta \rho}{\dot{\rho}}
    \\ & \mbox{}
        - \frac{1}{\dot{\rho}}
            \frac{\partial_i \delta \rho}{\dot{\rho}}
            \frac{\partial_j \delta \rho}{\dot{\rho}}
            \Big(
                T^{ij} + \rho g^{ij}
            \Big)
\end{split}
\end{equation}
In this expression, all perturbative quantities on the right-hand
side are evaluated on spatially flat slices.
After substitution in~\eqref{eq:psi-xfm} with $\xi^j = 0$, we find
\begin{equation}
\begin{split}
    \zeta = \mbox{} &
        - H \frac{\delta\rho}{\dot{\rho}}
        + H \frac{\delta\dot{\rho}}{\dot{\rho}} \frac{\delta\rho}{\dot{\rho}}
        - \frac{H}{2} \frac{\delta\rho}{\dot{\rho}}
            \frac{\partial}{\partial t} \frac{\delta\rho}{\dot{\rho}}
        - \frac{H}{2} \frac{\ddot{\rho}}{\dot{\rho}}
            \Big( \frac{\delta\rho}{\dot{\rho}} \Big)^2
        + \frac{2H}{\dot{\rho}} ( \delta T^{0i} + \rho \delta g^{0i} )
            \frac{\partial_i \delta \rho}{\dot{\rho}}
    \\ & \mbox{}
        - \frac{H}{\dot{\rho}}
            \frac{\partial_i \delta\rho}{\dot{\rho}}
            \frac{\partial_j \delta\rho}{\dot{\rho}}
            \Big(
                T^{ij} + \rho g^{ij}
            \Big)
        - \frac{1}{6a^2}
            \frac{\partial_i \delta\rho}{\dot{\rho}}
            \frac{\partial_i \delta\rho}{\dot{\rho}}
        - \frac{1}{3\dot{\rho}} \partial_i \vartheta_1 \partial_i \delta\rho
        + \frac{H}{4}
            \Big(
                \frac{\partial}{\partial t} \frac{\delta\rho}{\dot{\rho}}
            \Big)^2
        + \frac{\dot{H}}{2} \frac{(\delta \rho)^2}{\dot{\rho}} .
\end{split}
\label{eq:zeta-rho}
\end{equation}
Eq.~\eqref{eq:zeta-rho} is one of our central results.
It gives the curvature perturbation on uniform-density slices
in terms of the flat-gauge density perturbation,
the $0i$ components of the flat-gauge energy--momentum tensor
and metric,
and the scalar part of the flat-gauge shift vector encoded in
$\vartheta_1$.
It applies for any matter content.

For applications to inflation the matter theory
is given by an arbitrary number of scalar fields
interacting via a potential $V$.
The energy--momentum
tensor is
\begin{equation}
    T_{\mu\nu} =
        \partial_\mu \phi \partial_\nu \phi
        - \frac{1}{2} g_{\mu\nu} \partial^\lambda \phi \partial_\lambda \phi
        - g_{\mu\nu} V .
\end{equation}
It gives a background density $\rho = \dot{\phi}^2/2 + V$.
The density perturbation on flat slices is
\begin{equation}
\begin{split}
    \delta\rho = \mbox{} &
        - \alpha_1 \dot{\phi}^2
        + \dot{\phi}^\alpha \delta\dot{\phi}_\alpha
        + V_\alpha \delta\phi^\alpha
        - \dot{\phi}^\alpha \partial_i \vartheta_1 \partial_i \delta\phi_\alpha
        + \frac{1}{2} \delta\dot{\phi}^\alpha \delta\dot{\phi}_\alpha
        - 2 \alpha_1 \dot{\phi}^\alpha \delta\dot{\phi}_\alpha
    \\ & \mbox{}
        + \frac{\dot{\phi}^2}{2} (3\alpha_1^2 - 2\alpha_2)
        + \frac{1}{2} V_{\alpha\beta} \delta\phi^\alpha \delta\phi^\beta
        + \frac{1}{2a^2} \partial_i \delta\phi^\alpha \partial_i \delta\phi_\alpha ,
\end{split}
\end{equation}
and the $0i$ component is
\begin{equation}
    \delta T^{0i} = \frac{1}{a^2} \dot{\phi}^\alpha \partial_i \delta\phi_\alpha .
\end{equation}

\para{Explicit expressions}%
We can now give explicit expressions for the first- and second-order
components of $\zeta$.
We define these to satisfy $\zeta = \zeta_1 + \zeta_2 + \cdots$, and
as above $\zeta_n$ contains terms with exactly $n$ powers of the
field perturbations.
Dropping terms which decay when all wavenumbers correspond to
superhorizon scales, we find
\begin{subequations}
\begin{align}
    \zeta_1 = \mbox{} & \frac{1}{6\Mp^2 H^2 \epsilon}
        \Big(
            \dot{\phi}^\alpha \delta\dot{\phi}_\alpha
            + V_\alpha \delta\phi^\alpha
            - 2 \Mp^2 H^2 \epsilon \alpha_1
        \Big)
    \label{eq:zeta-one}
    \\
    \zeta_2 = \mbox{} & \frac{1}{6\Mp^2 H^2 \epsilon}
        \bigg(
            \frac{1}{2} \delta\dot{\phi}^\alpha \delta\dot{\phi}_\alpha
            + \frac{1}{2} V_{\alpha\beta} \delta\phi^\alpha \delta\phi^\beta
            - 2 \alpha_1 \dot{\phi}^\alpha \delta\dot{\phi}_\alpha
            + H^2 \Mp^2 \epsilon ( 3\alpha_1^2 - 2\alpha_2 )
    \nonumber
    \\ & \mbox{}
        \qquad\qquad\quad
            + H^2 \zeta_1
                \bigg[
                    \epsilon
                        \Big(
                            \frac{3+\epsilon}{H} \dot{\phi}^\alpha
                            - \frac{V_\alpha}{H^2}
                        \Big) \delta\phi^\alpha
                    - \frac{6+\epsilon}{H^2} \dot{\phi}^\alpha \delta\dot{\phi}_\alpha
                \bigg]
    \nonumber
    \\ &
        \qquad\qquad\quad
        \mbox{}
            + 3H^2 \zeta_1^2
                \Big[
                    6 \Mp^2 \epsilon + \frac{V_\alpha \dot{\phi}^\alpha}{H^3}
                \Big]
        \bigg) .
    \label{eq:zeta-two}
\end{align}
\end{subequations}
These expressions are exact, except for the neglect of decaying terms.
In deriving them we have made no use of the slow-roll approximation.

Eq.~\eqref{eq:zeta-two} shows that,
when derived using this method, the second-order
curvature perturbation
contains $\alpha_2$ and therefore apparently
depends on the nonlocal combination which appears in~\eqref{eq:lapse-two}.
If true this would be perplexing.
The explicit single-field expression given by Maldacena
contains no such terms~\cite{Maldacena:2002vr}.
The resolution is that,
in Maldacena's calculation,
the second-order lapse was removed
entirely by the Hamiltonian
constraint~\eqref{eq:hamiltonian-constraint}.

The existence of constraints means that
Eqs.~\eqref{eq:zeta-one}--\eqref{eq:zeta-two}
can be written in a number of superficially different ways.
One reason for doing so is that, because these rewritten formulations
contain different terms, their numerical properties can
differ even though they are mathematically equivalent.
If we choose to exploit
this freedom, however, we must remember that the Hamiltonian
constraint mixes terms of different orders in the field
fluctuations $\delta\phi^\alpha$.
Therefore, in quantities which depend on both $\zeta_1$ and $\zeta_2$,
we must use expressions which have been simplified in the same way.
Failure to do so will lead to a mismatch.
In particular this applies when computing the three-point
function $\langle \zeta(\vect{k}_1) \zeta(\vect{k}_2) \zeta(\vect{k}_3)
\rangle$
from $n$-point functions of the field fluctuations.

One option is to remove $\alpha_2$ entirely.
This will leave a purely local expression comparable to the one
obtained by Maldacena.
This choice gives
\begin{subequations}
\begin{align}
    \zeta_1^{\text{local}} = \mbox{} &
        \frac{1}{2 H^2 \Mp^2 \epsilon(3-\epsilon)}
        \Big(
            \dot{\phi}^\alpha \delta\dot{\phi}_\alpha
            + V_\alpha \delta \phi^\alpha
        \Big)
    \label{eq:zeta-local-one}
    \\
    \zeta_2^{\text{local}} = \mbox{} &
        \frac{1}{2 H^2 \Mp^2 \epsilon(3-\epsilon)}
        \bigg(
            \Big[
                \frac{1}{2} V_{\alpha\beta}
                + \frac{\dot{\phi}_\alpha \dot{\phi}_\beta}{\Mp^2}
                    \Big(
                        \frac{9}{2\epsilon}
                        - \frac{9}{2}
                        + \epsilon
                        + \frac{3-\epsilon}{4\epsilon^2}
                            \frac{V_\gamma \dot{\phi}^\gamma}{H^3 \Mp^2}
                    \Big)
            \Big] \delta \phi^\alpha \delta\phi^\beta
    \nonumber
    \\ &
        \qquad\qquad\qquad\qquad
        \mbox{}
            + \frac{\dot{\phi}_\alpha \dot{\phi}_\beta}{H\Mp^2}
                \Big(
                    \frac{3}{\epsilon}
                    - 2
                \Big) \delta\phi^\alpha \delta\dot{\phi}^\beta
            + \frac{1}{2} \delta\dot{\phi}^\alpha \delta\dot{\phi}_\alpha
        \bigg) .
    \label{eq:zeta-local-two}
\end{align}
\end{subequations}
Different forms for $\zeta_2$
can be obtained by further use of the first-order
Hamiltonian constraint.
For example, the cross-term $\delta\phi^\alpha \delta\dot{\phi}^\beta$
could be eliminated entirely at the expense of a more complex
coefficient for the $\delta\phi^\alpha \delta\phi^\beta$
term.

If we are prepared to tolerate residual nonlocal
terms, we could alternatively
use the Hamiltonian constraint to
simplify $\zeta_1$ and $\zeta_2$ as much as possible.
One choice is
\begin{subequations}
\begin{align}
    \zeta_1^{\text{simple}} = \mbox{} &
        - \frac{\dot{\phi}^\alpha \delta\phi_\alpha}{2H \Mp^2 \epsilon}
    \label{eq:zeta-simple-one}
    \\
    \zeta_2^{\text{simple}} = \mbox{} &
        \frac{1}{6H^2 \Mp^2 \epsilon}
        \bigg(
            \frac{\dot{\phi}_\alpha \dot{\phi}_\beta}{\Mp^2}
                \Big[
                    - \frac{3}{2}
                    + \frac{9}{2\epsilon}
                    + \frac{3}{4\epsilon^2} \frac{V_\gamma \dot{\phi}^\gamma}{\Mp^2 H^3}
                \Big] \delta\phi^\alpha \delta\phi^\beta
            + \frac{3}{H \epsilon}
                \frac{\dot{\phi}_\alpha \dot{\phi}_\beta}{\Mp^2}
                \delta\phi^\alpha \delta\dot{\phi}^\beta
    \nonumber
    \\ &
        \qquad\qquad\quad
        \mbox{}
        - 3 H \partial^{-2}
            \Big[
                \partial_j \delta\dot{\phi}^\alpha \partial_j \delta\phi_\alpha
                + \delta\dot{\phi}^\alpha \partial^2 \delta\phi_\alpha
            \Big]
        \bigg) .
    \label{eq:zeta-simple-two}
\end{align}
\end{subequations}
As above
the $\delta\phi^\alpha \delta\dot{\phi}^\beta$ terms can be removed,
if desired,
using the first-order constraint.
This form of $\zeta_1$ is especially simple,
being the multiple-field generalization
of the estimate
$\zeta \sim H \delta\phi / \dot{\phi}$
obtained in early
calculations~\cite{Guth:1982ec,Bardeen:1983qw,Lyth:1984gv}.
It coincides with the first-order expression
obtained by direct calculation of
the
comoving-gauge curvature perturbation $\R$,
and therefore
reproduces the first-order relation $\zeta_1 = \R_1$
on superhorizon scales
which was discussed in~\S\ref{sec:gauges}.
Because it requires the constraint equations
this relationship is a consequence of Einstein
gravity and need not hold more generally.

Eqs.~\eqref{eq:zeta-local-one}--\eqref{eq:zeta-local-two}
and~\eqref{eq:zeta-simple-one}--\eqref{eq:zeta-simple-two}
are exactly equivalent.
Neither involves any form of approximation except
that because we have neglected terms which decay when all
wavenumbers are associated with superhorizon scales
they are valid only in this limit.
Which we
use is a matter of our own convenience.
The only thing
we cannot do is mix (for example)
the simple first-order expression $\zeta_1^{\text{simple}}$ with the
local second-order result $\zeta_2^{\text{local}}$,
or vice-versa.

Which set is most convenient will depend on the problem at hand.
Eq.~\eqref{eq:zeta-simple-two} shows that
it is possible to compute the curvature perturbation
knowing only $\dot{\phi}^\alpha$, $H$ and $V_\alpha$
from the background, provided we are prepared to tolerate the
nonlocal terms.
In contrast with~\eqref{eq:zeta-local-two} it is not necessary
to know
the second derivative $V_{\alpha\beta}$
and we do not need a term quadratic in the derivatives
$\delta\dot{\phi}^\alpha$.
When used to obtain correlation functions of $\zeta$
this last property reduces the number of $n$-point functions of
the fields and their derivatives which must be computed.

With the guarantee provided
by~\eqref{eq:zeta-local-one}--\eqref{eq:zeta-local-two}
that it is \emph{possible} to write a purely local formula for $\zeta$,
the nonlocal terms in~\eqref{eq:zeta-simple-two} are harmless.
For computations of $n$-point functions, which naturally take place
in Fourier space,
they merely become constant factors of $k$.
Our numerical experiments suggest that
Eqs.~\eqref{eq:zeta-simple-one}--\eqref{eq:zeta-simple-two}
may even be preferable
to~\eqref{eq:zeta-local-one}--\eqref{eq:zeta-local-two}
because there are fewer cancellations between large contributions.
This is especially noticeable in models where $\zeta$ is conserved
at or after the end of inflation.
Conservation relies on a delicate interplay between
separate terms in $\zeta$ which may themselves be varying quite
rapidly.

\section{Comparison with the separate universe picture}
\label{sec:separate-universe}

The flat-gauge results for $\vartheta$
quoted in Eqs.~\eqref{eq:hamiltonian-first}--\eqref{eq:hamiltonian-second}
show that---up to second order
in fluctuations, and in coordinates where the spatial metric is
diagonal---the shift vector
$N^j$ approaches zero on superhorizon
scales~\cite{Weinberg:2008nf,Weinberg:2008si,Sugiyama:2012tj}.
In these coordinates the
only surviving
perturbation to the metric on superhorizon scales
is the lapse $\alpha$ which can be absorbed
into a shift of time.

After making this shift the metric is unperturbed. Therefore
the
equations for each matter species must be those of the
homogeneous, unperturbed universe, up to corrections of order
$(k/aH)^2$,
except with initial conditions displaced by the time shift
necessary to remove $\alpha$.
When promoted to
all orders in fluctuations this argument constitutes the
separate universe approach~\cite{Starobinsky:1986fxa,Sasaki:1995aw,Wands:2000dp,
Rigopoulos:2003ak,Lyth:2004gb,Lyth:2005fi}.
The necessary decay of the shift vector $N^j$
on superhorizon scales to all orders in perturbation theory
was shown by
Weinberg~\cite{Weinberg:2008nf,Weinberg:2008si}
and later strengthened by
Sugiyama, Futamase \& Komatsu~\cite{Sugiyama:2012tj}.
The conclusion is that superhorizon-sized regions
evolve individually like an unperturbed universe.

This formalism can be used to study the behaviour
of superhorizon-scale perturbations by comparing
the behaviour of each quantity of interest
on fixed
spatial hypersurfaces drawn from our choice of slicing.
To do so we must know how the background solutions,
parametrized in terms of this slicing,
change under a shift
of their initial conditions~\cite{Seery:2012vj}.
Therefore, in the separate universe approach,
choice of gauge is encoded
as the choice of time variable~\cite{Sasaki:1998ug}.

\para{Gauge transformations in the separate universe approach}
In this section we use the separate universe
approach to compute the gauge transformation
between $\delta\phi^\alpha$ and $\zeta$.
Versions of this calculation have been given before.
Anderson et al. collected formulae valid to third-order
on superhorizon scales, invoking the slow-roll expansion~\cite{Anderson:2012em}.
A derivation of the second-order gauge transformation
was given in Ref.~\cite{Seery:2012vj} using purely geometrical
methods on the phase space of solutions to the background equations.

The flat slicing corresponds to hypersurfaces separated by
equal amounts of expansion $N$, where $N(t_1, t_2) = \ln a(t_2)/a(t_1)$
measures the growth of the scale factor between
times $t_1$ and $t_2$.
The uniform-density slicing corresponds to hypersurfaces separated
by equal intervals of $\rho$.
In the separate universe approach, changing gauge from
the flat to uniform density slicings corresponds to changing
time variable from $N$ to $\rho$.

Consider an initial spatially flat hypersurface on which the
density can be written $\rho(\phi^\alpha, \dot{\phi}^\alpha)$.
Define some fixed value $\rho_\ast$ which is smaller than
$\rho$ everywhere on the hypersurface of interest,
and write $\Delta \rho = \rho_\ast - \rho$.
At each point $p$ on the hypersurface
we evolve the background equations of motion
(with initial conditions taken from their
values at $p$)
until the density reaches the constant value $\rho_\ast$,
and record the expansion $\Delta N$ which is accumulated.
Because $\rho$ varies over the slice $\Delta N$
will vary from point to point.
Its variation $\delta(\Delta N)$ represents
a modulation $\det h \sim \e{6 \delta(\Delta N)}$
of the proper volume on the final slice of fixed
density, and therefore
we can identify $\zeta = \delta(\Delta N)$.

\para{Uniform-density gauge curvature perturbation}%
If $\Delta\rho$ is not too large the expansion
accumulated during this evolution can be written
\begin{equation}
    \Delta N(p) =
        \left.\frac{\d N}{\d \rho}\right|_p \Delta \rho|_p
        + \frac{1}{2} \left.\frac{\d^2 N}{\d \rho^2}\right|_p \big( \Delta \rho|_p \big)^2
        + \cdots .
\end{equation}
It varies over the initial slice
because each term is a function of position $p$.
If the variation $\delta\rho$ under changes of $p$
is also not too large, then the variation in
$\Delta N$
under a change of initial location is
\begin{equation}
    \zeta
    =
    \delta (\Delta N)
    =
        - \left.\frac{\d N}{\d \rho}\right|_p \delta \rho
        - \delta
            \left(
                \frac{\d N}{\d \rho}
            \right)
            \delta\rho
        + \frac{1}{2}
            \left. \frac{\d^2 N}{\d \rho^2} \right|_p \delta \rho^2
    \label{eq:separate-universe-zeta}
\end{equation}
Because our interest lies in the gauge transformation
at a fixed time we have neglected terms which vanish
in the limit $\Delta \rho \rightarrow 0$,
which corresponds to coincidence of the initial and final
slices.

To obtain explicit expressions we require the derivatives
\begin{subequations}
\begin{align}
    \frac{\d N}{\d \rho} & = - \frac{1}{6 \Mp^2 H^2 \epsilon} , \\
    \frac{\d^2 N}{\d \rho^2} & = \frac{1}{(6 \Mp^2 H^2 \epsilon)^2}
        \Big(
            2\epsilon - \frac{\dot{\epsilon}}{H\epsilon}
        \Big) .
\end{align}
\end{subequations}
Up to this point our expressions apply for an arbitrary matter
theory.
Specializing to the case of canonical scalar fields appropriate
for inflation,
the variation $\delta ( \d N / \d \rho)$ satisfies
\begin{equation}
    \delta\left( \frac{\d N}{\d \rho} \right)
        =
        \delta\phi^\alpha
        \frac{\partial}{\partial \phi^\alpha} \frac{\d N}{\d \rho}
        +
        \delta\dot{\phi}^\alpha
        \frac{\partial}{\partial \dot{\phi}^\alpha} \frac{\d N}{\d \rho}
        + \cdots .
\end{equation}
We also have the exact expression $\rho = V/(1-\epsilon/3)$,
from which the variation $\delta\rho$ can be computed.
The result is
\begin{subequations}
\begin{align}
    \zeta_1^{\delta N} & \mbox{} =
        \frac{1}{2\Mp^2 H^2 \epsilon (3-\epsilon)}
        \Big(
            \dot{\phi}^\alpha \delta \dot{\phi}_\alpha
            + V_\alpha \delta\phi^\alpha
        \Big)
    \\
    \zeta_2^{\delta N} & \mbox{} =
        \frac{1}{2\Mp^2 H^2 \epsilon (3-\epsilon)}
        \bigg(
            \frac{1}{2}
            \Big[
                V_{\alpha\beta} -
                \frac{V_\alpha V_\beta}{H^2 \Mp^2(3-\epsilon)}
                \Big(
                    1+ \frac{\dot{\epsilon}}{2H\epsilon^2}
                \Big) 
            \Big]
            \delta \phi^\alpha \delta\phi^\beta
    \nonumber
    \\ &
    \qquad\qquad\qquad\qquad\quad\;\;
    \mbox{}
            -
            \frac{\dot{\phi}_\alpha V_\beta + \dot{\phi}_\beta V_\alpha}
                {4H^2\Mp^2 \epsilon(3-\epsilon)}
            \Big(
                3 - \epsilon + \frac{\dot{\epsilon}}{2H \epsilon}
            \Big)
            \delta \phi^\alpha \delta\dot{\phi}^\beta
    \nonumber
    \\ &
    \qquad\qquad\qquad\qquad\quad\;\;
    \mbox{}
            +
            \frac{1}{2}
            \Big[
                \delta_{\alpha\beta}
                -
                \frac{\dot{\phi}^\alpha \dot{\phi}^\beta}{\Mp^2 H^2 \epsilon(3-\epsilon)}
                \Big(
                   6 - 3\epsilon + \frac{\dot{\epsilon}}{2H\epsilon}
                \Big)
            \Big]
            \delta\dot{\phi}^\alpha \delta\dot{\phi}^\beta
        \bigg) .
\end{align}
\end{subequations}
The first-order term $\zeta_1^{\delta N}$ agrees immediately with the
local expression $\zeta_1^{\text{local}}$ given in Eq.~\eqref{eq:zeta-local-one}.
Although $\zeta_2^{\delta N}$ is superficially different
to $\zeta_2^{\text{local}}$ they can be made to agree using the
first-order Hamiltonian constraint and the equation of motion
for the background scalar field.
This gives an explicit demonstration
(assuming Einstein gravity)
that the gauge transformation
derived from the separate universe approach agrees with
the one derived from traditional cosmological perturbation theory.
In practice, if a local expression is required, the more compact
form~\eqref{eq:zeta-local-one}--\eqref{eq:zeta-local-two}
is likely to be preferable.

\section{Conclusions}
\label{sec:conclusions}

In this paper
we give a formula for the
uniform-density gauge curvature perturbation
written explicitly in terms of the scalar field fluctuation $\delta\phi^\alpha$
defined on spatially-flat slices.
This formula is needed to compute
observable quantities from second-order perturbation theory,
including the bispectrum
$\langle \zeta(\vect{k}_1) \zeta(\vect{k}_2) \zeta(\vect{k}_3) \rangle$.

Our results can be written in different ways using the Hamiltonian constraint.
In particular, although the expressions obtained directly from
cosmological perturbation theory involve `nonlocal' terms
which depend on the inverse Laplacian $\partial^{-2}$---and are therefore
na\"{\i}vely incompatible with the separate universe approach---we
show that that these terms can be removed using the constraints.
After doing so the results of perturbation theory and the
separate universe approach agree.
Our final results, especially
Eqs.~\eqref{eq:zeta-simple-one}--\eqref{eq:zeta-simple-two}
are compact, simple
and can be used directly in numerical calculations.
We have tested their validity using integrations of
the two- and three-point functions
$\langle \delta\phi^\alpha(\vect{k}_1) \delta\phi^\beta(\vect{k}_2) \rangle$
and
$\langle \delta\phi^\alpha(\vect{k}_1) \delta\phi^\beta(\vect{k}_2)
\delta\phi^\gamma(\vect{k}_3) \rangle$.
Using these gauge transformations
we confirm the expected behaviour of
$\langle \zeta(\vect{k}_1) \zeta(\vect{k}_2) \rangle$
and
$\langle \zeta(\vect{k}_1) \zeta(\vect{k}_2) \zeta(\vect{k}_3) \rangle$,
including accurate
conservation when all isocurvature modes become quenched.

\para{Comparison with Christopherson et al}%
While this paper was in preparation,
a preprint was released by
Christopherson, Nalson \& Malik which also gives
an explicit expression for $\zeta$
in terms of $\delta\phi^\alpha$ up to second order~\cite{Christopherson:2014bea}.

To aid comparison, we briefly list the similarities and
differences between our calculations.
First, Christopherson et al. adopt a different definition
of density. Our definition, $T_{ab} n^a n^b$,
gives
$\rho = \pi^2/2N^2 + (\partial \phi)^2/2a^2 + V$
expressed in
coordinates adapted to the slicing,
where $\pi^\alpha \equiv \dot{\phi}^\alpha - N^m \partial_m \phi^\alpha$.
It corresponds to what Hwang \& Noh called
the normal frame~\cite{Noh:2003yg}.
Christopherson et al. define the density in what
Hwang \& Noh call the energy frame, giving
$\rho = \pi^2/2N^2 - (\partial \phi)^2/2a^2 + V$,
again in coordinates adapted to the slicing.
When all wavenumbers correspond to superhorizon scales
the spatial gradients decay and these expressions agree.
Therefore, under the same circumstances,
our definitions of the uniform density slicing
will also agree.

Second, our definitions of the curvature perturbation
are different. Christopherson et al. adopt the definition of
Malik \& Wands~\cite{Malik:2008im},
in which the spatial metric is written
(including all orders in perturbation theory)
\begin{equation}
    h_{ij} = a^2
        \Big[
            ( 1 - 2 \psi_\text{MW} ) \delta_{ij}
            + \partial_j F_{i} + \partial_i F_{j}
            + \partial_i \partial_j E
            + \frac{1}{2} h_{ij}
        \Big]
        \d x^i \d x^j .
\end{equation}
where $F_i$ is divergenceless, and $h_{ij}$ is transverse
and tracefree.
Malik \& Wands define the curvature perturbation to be $\psi_\text{MW}$.
Our definition is $\psi = (1/6)\ln \det (h_{ij}/a^2)$,
because it is this quantity which is known to be conserved
on superhorizon
scales~\cite{Rigopoulos:2003ak,Malik:2005cy,Assassi:2012et}.
The Malik--Wands definition $\psi_{\text{MW}}$
is not equivalent to the determinant of $h_{ij}$
unless $E = F_i = h_{ij} = 0$.
In that case
the first-order parts of $\psi$ and $\psi_{\text{MW}}$
agree,
and the second-order parts are related by
$\psi_{\text{MW}|2} = \psi_2 + 2 (\psi_1)^2$~\cite{Lyth:2005du}.

Finally, we simplify our expressions using the
Hamiltonian constraint,
which Christopherson et al.
refer to as the momentum equation.
Christopherson et al. work only with cosmological
perturbation theory, not the separate universe
approach, and do not eliminate the nonlocal
terms which appear in their expressions.

\begin{acknowledgments}
We would like to thank Karim Malik for helpful comments on
a draft version of this paper.
DS acknowledges support from the Science and Technology
Facilities Council [grant number ST/L000652/1].
DS and JE acknowledge support from the Leverhulme Trust.
The research leading to these results has received funding from
the European Research Council under the European Union's
Seventh Framework Programme (FP/2007--2013) / ERC Grant
Agreement No. [308082]. DM is supported by a Royal Society University 
Research Fellowship, and was supported 
by the Science and Technology Facilities
Council during the majority of this work [grant number ST/J001546/1]. JF is supported by IKERBASQUE, the Basque Foundation for Science.
\end{acknowledgments}

\newpage
\bibliography{bibliography}
\end{document}